# From Historical Puzzles to Grammatical Constraints: Circular Partitions, Generalized Run-Length Encodings, and Polynomial-Time Decidability

Omid Khormali[1,*] Ghaya Mtimet[1] Nuh Aydin[2]

[1]Department of Mathematics, University of Evansville, Evansville, IN 47722, USA
Emails: ok16@evansville.edu, gm163@evansville.edu

[2]Department of Mathematics and Statistics, Kenyon College, Gambier, OH 43022, USA
Email: aydinn@kenyon.edu

**Abstract**

Motivated by a historical combinatorial problem that resembles the well-known Josephus problem, we investigate circular partitions, formulate problems in deterministic finite automata, and develop algorithms to solve problems in this context. The historical problem involves arranging individuals on a circle and eliminating every $k$-th person until a desired group remains. We analyze both removal and non-removal approaches to circular partitioning, establish conditions for balanced partitions, and present explicit algorithms. We introduce generalized run-length encodings over partitioned alphabets to capture alternating letter patterns and compute their cardinalities using Stirling numbers of the second kind. Connecting these combinatorial structures to formal language theory, we formulate an existence problem: given a context-free grammar over a dictionary and block-pattern constraints on letters, does a valid sentence exist? We prove decidability in polynomial time by showing block languages are regular and applying standard parsing techniques. Complete algorithms with complexity analysis are provided and validated through implementation on both historical and synthetic instances.

**Keywords:** Deterministic finite automata, Josephus problem, circular partitions, context-free grammars, formal languages, decidability, block languages.

**2020 Mathematics Subject Classification:** 68Q45, 68R15, 68W40

## 1 Introduction

This paper is inspired by a clever historical strategy to solve a interesting problem that appears in the Islamic narrative tradition. A written version of the story is found in *Hāshiyat al-Aṭṭār alá Jam*

*Corresponding author: ok16@evansville.edu

*al-Jawāmi li Ibn al-Subkī*, a commentary by the Egyptian scholar al-Aṭṭār [1], who lived in the early 19th century (d. 1250 AH/1835 CE). [1] Beyond this textual source, the story is also known among Islamic scholars and preachers, who have referred to it in oral contexts and attributed it to early Islamic times.

In this account, a group of 30 individuals, comprising 15 believers and 15 disbelievers, was aboard a ship in crisis, and it was determined that half of the passengers needed to be removed to save the vessel. The need to remove a certain number of passengers from a vessel that is in danger of sinking seems to be an old and common story that existed in various traditions before Islam. The group turned to one man for a fair method to decide who should remain. Seeking to ensure the safety of the believers, the man devised a symbolic strategy based on the structure of the Arabic alphabet and a carefully chosen sentence. He began by dividing the Arabic letters into two equal groups, those with dots and those without, and created a sentence consisting of exactly 30 letters, evenly split between the two categories. The passengers were arranged in a circle, and each was assigned a letter from the sentence in order. The letters corresponding to believers were drawn from one group, and those for disbelievers from the other. Beginning with the first letter of the sentence, every ninth person was eliminated cyclically until only 15 remained. The goal was to ensure that all remaining individuals were from the believer group.

The poetic verse used in the version preserved in the book is [2]

الله يقضي بكل يسر    ويرزق الضيف حيث كان

While al-Aṭṭār's commentary provides an accessible written record of this mnemonic strategy, earlier references to similar mathematical puzzles appear in medieval Arabic biographical literature. The 14th-century scholar Ṣalāḥ al-Dīn al-Ṣafadī (d. 764 AH/1363 CE) documented this type of problem in his biographical dictionary *Ayān al-Aṣr wa-Awān al-Naṣr* [5]. Al-Ṣafadī reports that he heard the verse above (the first verse) and was inspired to create his own mnemonic solution for the same problem. His verse is: [3]

ولما فتنت بلحظ له    عذلت فما خفت من شامت

Al-Ṣafadī also composed several other mnemonic verses for this circle-based counting puzzle, which we will examine in Section 3. These 14th-century references establish that such counting problems and their mnemonic solutions were circulating among Arabic-speaking scholars at least five centuries before al-Aṭṭār's work, suggesting a longer tradition of recreational mathematics in Islamic intellectual culture. In Section 3, we examine al-Ṣafadī's mnemonic verses in detail and demonstrate how they encode the mathematical solution to the circle arrangement problem.

This strategic use of a sentence, combined with the rhythm of 9 and the binary alphabet partition, reveals an intriguing combination of mathematical patterns and language. The framework of this problem, involving the removal of every ninth individual from a circle of 30, closely resembles the classical Josephus problem [6] that predates Islamic sources and has been widely studied in combinatorics and algorithm design. However, the integration of letter assignment from an Arabic sentence, carefully constructed to control the outcome, appears unique to Islamic narrative or exegetical contexts.[4] While the mathematical parameters are similar, it remains an open question whether the numbers 30 and 9 were independently formulated within Islamic literature or reflect influence from earlier Josephus-type formulations. This is a subject better addressed by historians of Islamic science and intellectual transmission.

[1] See *Hāshiyat al-Aṭṭār alá Jam al-Jawāmi*, vol. 2, Dār al-Kutub al-Ilmiyya, p. 264.

[2] This sentence may be translated to English as "God decrees all matters with ease and provides for the guest wherever he may be."

[3] This sentence may be translated as "And when I was captivated by a glance from him, I was reproached, yet I feared no gloater."

A strikingly similar version, featuring 15 Turks and 15 Christians aboard a ship, is found in James R. Newman's *The World of Mathematics* (1956) [3]. In this version, every ninth person is eliminated, with the goal of ensuring the Christians survive. In contrast, Newman's version lacks the linguistic encoding via a 30-letter Arabic sentence and the associated alphabetic partitioning present in the Islamic formulation. Although it is unclear whether Newman's version was derived from the Islamic narrative or developed independently, the textual record suggests that the Arabic version predates the appearance of Western print, and the relationship between the two versions remains unclear. Further historical research is needed to determine whether the version attributed to Newman has antecedents in earlier Western literature or reflects influence from Islamic sources.

In this paper, we analyze this culturally embedded elimination method through the lens of discrete mathematics. We explore its combinatorial properties, compare it with standard Josephus-type formulations, and discuss its implications for algorithmic fairness and group selection.

First, we examine the problem of partitioning $n$ objects into $r$ groups. We approach this problem by placing the objects on a circle and assigning group labels using a cyclic stepping procedure. Our aim is to investigate how this circular method can produce balanced and fair groupings. Throughout this paper, we define the step size $k$ in terms of the positions on the circle. Starting from an initial position (which we designate as position 1, not removed to begin with), we remove the object at position $k$, then at position $2k$, position $3k$, and so on, wrapping around the circle modulo $n$ (so after reaching position $n$, we continue to positions that exceed $n$ by taking their remainder when divided by $n$). Note that the distance from the starting position to the first removed object is $k-1$ positions, while the distance between any two consecutively removed objects is exactly $k$ positions. For clarity, when we refer to "step size $k$," we mean that removed objects are located at positions $k, 2k, 3k, \ldots$ relative to the starting position, computed modulo $n$ as we wrap around the circle.

## 2 The Partition Problem

Let $n$ be the number of objects to be partitioned into $r \geq 2$ groups. We arrange these objects randomly on a circle and label their positions. The labeling can proceed either clockwise or counterclockwise around the circle. We then apply a cyclic stepping procedure with a fixed step size $k \geq 1$, moving through the positions in the order of their labels. We consider two distinct approaches: the *non-removal approach* where visited objects remain on the circle and are simply marked as visited, and the *removal approach* where visited objects are removed from the circle

[4] See `https://niefrar.org/?p=74` for a modern description of the story.

after assignment, similar to the classical Josephus problem.

In the non-removal approach, traditional balanced partitioning uses modular arithmetic with step-size $k = r$ and the first object in each round always marked. However, many applications require different assignment orders while maintaining balance. Our step-size parameter $k$ generalizes this approach, where gcd relationships with $r$ characterize feasible values of $k$. While final group sizes remain balanced regardless of $k$, the sequence of object assignment varies, enabling optimization for locality, fault tolerance, and access patterns.

### 2.1 Partition Algorithms

In the following theorem, we present algorithms to solve this problem for both the removal and non-removal approaches.

**Theorem 2.1.** *For any positive integers $n, r$ and $k$ such that $r \leq n$, there exists a balanced partition of the $n$-set $[n] = \{1, 2, \ldots, n\}$ into $r$ groups using step size $k$.*

This theorem can be proved using a removal or non-removal approach. They may create distinct partitions, but either approach accomplishes the goal. We present a solution for each case separately.

**Remark 1.** *When $k > n$, stepping by $k$ positions around a circle of $n$ objects is equivalent to stepping by $k \bmod n$ positions. Thus, without loss of generality, we may assume $k \leq n$ in the proofs, but the theorem holds for all positive integers $k$.*

*Proof 1: Removal Approach.* We provide a constructive proof using a Josephus-like elimination process. Let $s \geq 0$ be the smallest non-negative integer such that $r \mid (n+s)$. We can express $s = a \cdot \lceil n/r \rceil + b$ where $a \geq 0$ and $0 \leq b < \lceil n/r \rceil$. We introduce $s$ dummy objects $\{s_1, s_2, \ldots, s_{a \cdot \lceil n/r \rceil + b}\}$, and proceed as follows:

1. Set $N = n + s$ as the total number of objects and $R = r + a$ as the total number of groups.
2. Place all $N$ objects around a circle, labeled 1 to $N$, where the $n$ real objects occupy positions 1 to $n$, and the $s$ dummy objects occupy positions $n + 1$ to $N$.
3. Starting from position 1, select objects using step size $k$ and remove each selected object. (That is, position $k$ is selected first, then position $2k$, and so on, with position 1 serving as the reference point.) During selection:
   - Assign objects sequentially to groups such that first $N/R = \lceil n/r \rceil$ objects for the first group, and the second $N/R$ objects for the second group, and so on. Then each group receives exactly $N/R$ objects.
   - Maintain a counter $i$ for dummy objects encountered. Set the initial value $i = 0$.
   - When encountering a dummy object, we do
     - Increment counter, $i \leftarrow i + 1$
     - If $i \leq b$, this dummy object must go to group $i$.
       * If currently filling group $i$, then assign the dummy object directly to group $i$

* If currently filling group $j \neq i$, then:
      - If group $i$ is not yet filled, then place the new dummy in group $i$ and continue filling group $j$ with real objects
      - If group $i$ is already filled with real objects, then swap the dummy object with a real object from group $i$
  - If $i > b$, then this dummy object goes to the one of the dummy groups ($r + 1$ through $R$) and is normally assigned to the current group being filled
- This process ensures:
  - The 1st dummy object encountered always ends up in group 1
  - The 2nd dummy object encountered always ends up in group 2
  - ... and so on up to the $b$-th dummy object in group $b$
  - Groups 1 through $b$ each contain exactly one dummy object
  - All remaining dummy objects go to groups $r + 1$ through $R$

Continue until all objects are assigned to groups $1, 2, \ldots, R$.

4. Remove the $b$ dummy objects from groups 1 through $b$.

5. Discard the $a$ dummy groups (groups $r + 1$ through $R$) along with their dummy objects.

Note that since $s$ is the smallest non-negative integer satisfying $r \mid (n + s)$, we have $s < r$. Consequently, $b \leq s < r$, ensuring that the first $b$ dummy objects can be distributed to $b$ distinct real groups.

Now we show that the algorithm generates a balanced partition. After step 3, we have:

- Groups 1 through $b$: Each has $\lceil n/r \rceil$ objects (including one dummy object)
- Groups $b + 1$ through $r$: Each has $\lceil n/r \rceil$ real objects
- Groups $r + 1$ through $R$: Each has $\lceil n/r \rceil$ dummy objects

After removing dummy objects and dummy groups:

- Groups 1 through $b$: Each has $\lceil n/r \rceil - 1 = \lfloor n/r \rfloor$ real objects
- Groups $b + 1$ through $r$: Each has $\lceil n/r \rceil$ real objects

Since $n = r \cdot \lfloor n/r \rfloor + b$ (where $b = n \bmod r$), we need exactly $b$ groups with $\lceil n/r \rceil$ objects and $r - b$ groups with $\lfloor n/r \rfloor$ objects for a balanced partition. This is precisely what we achieve. □

**Remark 2.** *In the proof, the first $b$ dummy objects are assigned to the first $b$ real groups for convenience. However, this choice is arbitrary, and the first $b$ dummy objects may be assigned to any $b$ real groups without affecting the argument.*

*Proof 2: Non-Removal Approach.* We provide an alternative constructive proof where objects remain in fixed positions throughout the partitioning process. We first define two core algorithms that we will use in our case analysis.

Algorithm A:
Input: $m$ objects such that $\gcd(m, k) = g$.
**Note:** Unlike the removal approach where position 1 serves only as a reference point, in Algorithm A the starting position is assigned to a group. This difference does not affect the partition structure, as the algorithm partitions objects into cycles determined by $\gcd(m, k)$.

1. Place objects around a circle labeled $1$ to $m$.
2. Starting from position $1$, assign it to group $1$.
3. Step by $k$ positions, assigning visited objects to the same group until returning to start.
4. Repeat from the next unvisited position for the next group, continuing until all objects are grouped.

Algorithm B:
Input: $m$ objects such that $\gcd(m, k) = g$ and $\gcd(m/g, k) = f$.

1. Apply Algorithm A to create initial groups.
2. Place each group's objects on a separate circle.
3. Partition each group into subgroups using the specified step size $k$ using Algorithm A with $m/g$ objects.

Note that Algorithm A creates $g$ groups, each containing exactly $\frac{m}{g}$ objects. And Algorithm B creates $gf$ groups, each containing exactly $\frac{m}{gf}$ objects. Also, $m$, which is the number of objects, will be chosen appropriately for each case.

Now, we consider cases based on the relationship among $n$, $r$, and $k$, and in each case we choose $m$ appropriately.

Case 1: $k \geq r$
Case 1.1: $\gcd(k, r) = r$

- If $\gcd(n, k) = r$: Algorithm A with $m = n$ objects creates $r$ groups of size $n/r$ each. Since $r \mid n$, this is perfectly balanced.
- If $\gcd(n, k) \neq r$: After adding $s$ dummy objects where $s < r$, Algorithm A with $m = n + s$ objects creates $r$ groups of size $(n + s)/r$. The dummy objects are distributed among $s$ groups, one per group. Note that in the worst case, if any group contains more than one dummy object, we can swap a dummy object with a real object from a group that contains no dummy object. After performing these swaps, we obtain $s$ groups, each containing exactly one dummy object. Removing them leaves $s$ groups with $(n + s)/r - 1$ objects and $r - s$ groups with $(n + s)/r$ objects, achieving balance.

Case 1.2: $\gcd(k, r) \neq r$

- Introduce $t$ dummy groups, each containing $\lceil n/r \rceil$ dummy objects, where

$$t = \text{the smallest integer such that } \gcd(k, r+t) = r+t.$$

- The total number of dummy objects is $t \cdot \lceil n/r \rceil$, so the new total number of objects is $N = n + t\lceil n/r \rceil$ with $r + t$ groups and the same step size $k$.

- As in Case 1.1, we now consider the two subcases

$$\gcd(N, k) = r + t \qquad \text{or} \qquad \gcd(N, k) \neq r + t.$$

- If $\gcd(N, k) = r + t$, then Algorithm A applied to the $m = N$ objects produces $r + t$ groups of equal size. Any real objects appearing in dummy groups are swapped with dummy objects appearing in real groups. After these swaps, all dummy groups contain only dummy objects, and all dummy objects are only in dummy groups.

- If $\gcd(N, k) \neq r+t$, we add $s$ dummy objects (with $1 \leq s \leq r+t-1$) so that $\gcd(N+s, k) = r + t$, and apply Algorithm A to the $m = N + s$ objects. This yields $r + t$ groups of size $\frac{N+s}{r+t}$.

  The total number of objects is $N + s = n + t\lceil n/r \rceil + s$. If all groups had size $\lceil n/r \rceil$, the total would be $(r + t)\lceil n/r \rceil$. The excess is

$$(n + t\lceil n/r \rceil + s) - (r + t)\lceil n/r \rceil = n + s - r\lceil n/r \rceil.$$

  When $r \nmid n$, let $b = n \bmod r$ where $1 \leq b < r$. Then $\lceil n/r \rceil = \lfloor n/r \rfloor + 1$, so

$$n - r\lceil n/r \rceil = n - r(\lfloor n/r \rfloor + 1) = (n - r\lfloor n/r \rfloor) - r = b - r.$$

  Thus the excess is $s + b - r$, which equals the number of groups with size $\lceil n/r \rceil + 1$.

  We designate $t$ dummy groups by selecting from groups with size $\lceil n/r \rceil + 1$ when available. If the number of such groups is less than $t$, we select all of them as dummy groups and choose the remaining from groups of size $\lceil n/r \rceil$.

  **Case (a):** If $s \leq r$, then there are $s + b - r < r$ groups of size $\lceil n/r \rceil + 1$. After placing dummy objects in the $t$ dummy groups, at most $s < r$ dummy objects remain for the $r$ real groups.

  **Case (b):** If $s > r$, then there are $s + b - r$ groups of size $\lceil n/r \rceil + 1$. We select all $t$ dummy groups from these larger groups. After placing $t(\lceil n/r \rceil + 1)$ dummy objects in dummy groups, the remaining dummy objects for real groups is

$$(t\lceil n/r \rceil + s) - t(\lceil n/r \rceil + 1) = s - t < r.$$

  In both cases, at most $r - 1$ dummy objects remain in real groups. As before, we swap real objects in dummy groups with dummy objects in real groups so that all dummy groups contain only dummy objects. Then we ensure each real group contains at most one dummy object by swapping excess dummy objects with real objects from groups containing no dummy objects.

- Finally, we remove all dummy groups and delete all dummy objects. Each real group then contains either $\lceil n/r \rceil$ or $\lfloor n/r \rfloor$ real objects.

Case 2: $k < r$
Case 2.1: $k > 1$ and $\gcd(r, k) = k$

- We add $s \geq 0$ dummy objects such that $\gcd(n + s, r) = k$.
- We use Algorithm B with $m = n + s$ objects. It first creates $k$ groups of size $(n + s)/k$ each.
- Each group is then partitioned into $r/k$ subgroups of size $(n + s)/r$.
- Since $s < r$, removing dummy objects (at most one per subgroup) yields balance.

Case 2.2: $k > 1$ and $\gcd(r, k) \neq k$

- Similar to Case 1.2, $t$ dummy groups are added so that $\gcd(r+t, k) = k$. Each dummy group contains $\lceil n/r \rceil$ dummy objects. Thus the total number of objects becomes $N = n+t\cdot \lceil n/r \rceil$.

  We again consider two subcases:

  – If $\gcd(N, r + t) = k$, then apply Algorithm B to the $m = N$ objects with $r + t$ groups and step-size $k$. Afterward, swap any real objects appearing in dummy groups with dummy objects in real groups so that all dummy groups contain only dummy objects. Then remove the dummy groups and dummy objects.
  – If $\gcd(N, r + t) \neq k$, add $s$ dummy objects so that $\gcd(N + s,\ r + t) = k$. Apply Algorithm B to the $m = N + s$ objects with $r + t$ groups and step-size $k$. Afterward, swap any real objects found in dummy groups with dummy objects originally assigned to dummy groups. For the real groups, ensure that at most one dummy object is in each real group. Finally, remove all dummy groups and dummy objects.

Case 2.3: $k = 1$

- Direct assignment of $\lceil n/r \rceil$ consecutive objects to each of the first $n \bmod r$ groups and $\lfloor n/r \rfloor$ objects to the remaining groups produces a balanced partition by definition.

In all cases, the final partition has each group containing either $\lfloor n/r \rfloor$ or $\lceil n/r \rceil$ objects, with exactly $n \bmod r$ groups having $\lceil n/r \rceil$ objects, thus achieving a balanced partition. □

**Remark 3.** *In the second proof, in Cases 1.1, 1.2, and 2.2, we did not explicitly state that, after ensuring all objects in dummy groups are dummy, the remaining dummy objects in real groups should be assigned to the real groups with smaller labels. We included only the condition that each real group has at most one dummy object after all swaps.*

## 2.2 Partition examples

Now we give some examples of the partition algorithms.

**Example 1.** *Removal Approach (Proof 1):*
*Let parameters be* $n = 10, r = 3, k = 4$. *We need to find the smallest* $s \geq 0$ *such that* $r \mid (n + s)$, *i.e.,* $3 \mid (10 + s)$. *So, we have* $s = 2$. *Also,* $s = a \cdot \lceil n/r \rceil + b$ *where* $\lceil 10/3 \rceil = 4$. *Then, we have* $a = 0$ *and* $b = 2$. *And* $N = n + s = 12$ *objects and* $R = r + a = 3$ *groups.*
*We place 10 real objects (positions 1 - 10) and 2 dummy objects (positions 11-12) around a circle. Starting from position 1 as the reference point with step size* $k = 4$, *we select and remove objects at positions* $4, 8, 12, \ldots$ *(modulo 12).*

| *Selection order* | *Position* | *Type* | *Assigned to* |
|---|---|---|---|
| *1* | *4* | *real* | *Group 1* |
| *2* | *8* | *real* | *Group 1* |
| *3* | *12* | *dummy* | *Group 1 (1st dummy* $\rightarrow$ *Group 1)* |
| *4* | *5* | *real* | *Group 1* |
| *5* | *10* | *real* | *Group 2* |
| *6* | *3* | *real* | *Group 2* |
| *7* | *11* | *dummy* | *Group 2 (1st dummy* $\rightarrow$ *Group 2)* |
| *8* | *7* | *real* | *Group 2* |
| *9* | *6* | *real* | *Group 3* |
| *10* | *9* | *real* | *Group 3* |
| *11* | *2* | *real* | *Group 3* |
| *12* | *1* | *real* | *Group 3* |

*After step 3 (before removing dummies), we have*

$$G_1 = \{4, 5, 8, 12\}, \quad G_2 = \{3, 7, 10, 11\}, \quad G_3 = \{1, 2, 6, 9\}.$$

*After removing the* $b = 2$ *dummy objects from Groups 1 and 2, we have*

$$G_1 = \{4, 5, 8\}, \quad G_2 = \{2, 7, 10\}, \quad G_3 = \{1, 2, 6, 9\}.$$

*This gives us one group of size 4 and two groups of size 3, achieving the balanced partition where* $b = n \bmod r = 1$ *group has* $\lceil 10/3 \rceil = 4$ *objects and* $r - b = 2$ *groups have* $\lfloor 10/3 \rfloor = 3$ *objects.*

**Example 2.** *Non-Removal Approach, Case 1.1:*
*Suppose the parameters are* $n = 10, r = 4, k = 4$. *Here* $k \geq r$, $\gcd(k, r) = \gcd(4, 4) = r$, *but* $\gcd(n, k) = \gcd(10, 4) = 2 \neq r$. *Thus, this is Case 1.1, the second sub-case.*
*We must add the smallest number* $s$ *of dummy objects such that* $\gcd(n + s, k) = r$. *Here,* $s = 2$ *since* $\gcd(12, 4) = 4 = r$. *So we work with* $m = N = 12$ *objects in Algorithm A and* $r = 4$ *groups. At the end, we remove the 2 dummy objects.*

| *Step* | *Start Pos* | *Visited sequence* | *Group* |
|---|---|---|---|
| *1* | *1* | *1, 5, 9* | *Group 1* |
| *2* | *2* | *2, 6, 10* | *Group 2* |
| *3* | *3* | *3, 7, 11(dummy)* | *Group 3* |
| *4* | *4* | *4, 8, 12(dummy)* | *Group 4* |

*Removing the two dummy objects, the final partition is*

$$G_1 = \{1, 5, 9\}, \quad G_2 = \{2, 6, 10\}, \quad G_3 = \{3, 7\}, \quad G_4 = \{4, 8\}.$$

**Example 3.** *Non-Removal Approach, Case 1.2:*
*Assume the parameters are* $n = 12, r = 3, k = 4$. *Here* $k \geq r$ *and* $\gcd(k, r) = \gcd(4, 3) = 1 \neq r$, *so we are in Case 1.2.*
*We add* $t = 1$ *dummy group, since* $\gcd(4, 4) = 4$. *Each dummy group receives* $\lceil 12/3 \rceil = 4$ *dummy objects. Thus* $N = 16$ *objects,* $r + t = 4$ *groups, step-size* $k = 4$.
*Applying Case 1.1 to* $(m = N = 16$ *in Algorithm A*$, r = 4, k = 4)$ *gives*

| *Step* | *Start Pos* | *Visited sequence* | *Group* |
|---|---|---|---|
| *1* | *1* | *1, 5, 9, 13(dummy)* | *Group 1* |
| *2* | *2* | *2, 6, 10, 14(dummy)* | *Group 2* |
| *3* | *3* | *3, 7, 11, 15(dummy)* | *Group 3* |
| *4* | *4* | *4, 8, 12, 16(dummy)* | *Group 4(dummy)* |

*Swapping step: If any real objects had been placed in the dummy group, we would swap them with dummy objects that landed in non-dummy groups. So the updated groups are*

| *Elements* | *Group* |
|---|---|
| *1, 5, 9, 4* | *Group 1* |
| *2, 6, 10, 8* | *Group 2* |
| *3, 7, 11, 12* | *Group 3* |
| *13(dummy), 14(dummy), 15(dummy), 16(dummy)* | *Group 4(dummy)* |

*Removal step: Remove Group 4 entirely (since it contains only dummy objects).*

*Final partition into* $r = 3$ *groups is*

$$G_1 = \{1, 4, 5, 9\}, \quad G_2 = \{2, 6, 8, 10\}, \quad G_3 = \{3, 7, 11, 12\}.$$

## 3 Historical Example and Alphabet-Based Partitioning

In this section, we demonstrate the well-known historical example of thirty individuals divided into two equal groups of fifteen. The individuals are arranged on a circle, each assigned to a letter drawn sequentially from an Arabic sentence of length 30. The elimination process removes every ninth person (equivalently, using step size $k = 9$), until exactly fifteen remain. To explain this demonstration, we first recall the Arabic alphabet and its classical partition into two sub-alphabets: letters written with dots and letters written without dots. This binary division plays a central role in the construction of the sentence and the assignment of passengers.

Arabic Alphabet Partition: The full Arabic alphabet (28 letters) can be divided into two subgroups as follows:

- *Dotless letters (14):* ا، ح، د، ر، س، ص، ط، ع، ك، ل، م، هـ، و، ى
- *Dotted letters (15):* ، ب، ت، ث، ج، خ، ذ، ز، ش، ض، ظ، غ، ف، ق، ن، ئ، ي

**Note:** The letter yā appears in multiple forms in Arabic orthography. The dotted ي (standard yā), dotted ئ (yā with hamza), and dotless ى (*alif maqṣūrah*). For ئ, Ryding [4, p. 16, section 3.3.1.1] notes that when yā serves as a seat (carrier) for hamza, it loses its two dots. However, since the underlying letter remains yā, which is inherently dotted, ئ is classified as a dotted letter based on its base letter.[5]

Figure 1 shows the 30 letters of the Arabic sentence below arranged in a circle: blue nodes are dotless letters, red nodes are dotted letters. Counting from the initial position marked in the figure, every ninth letter is eliminated. It turns out that the first 15 letters deleted in this way are all red (dotted) letters. The Arabic sentence, which was introduced earlier, and the configuration of the letters are as follows.

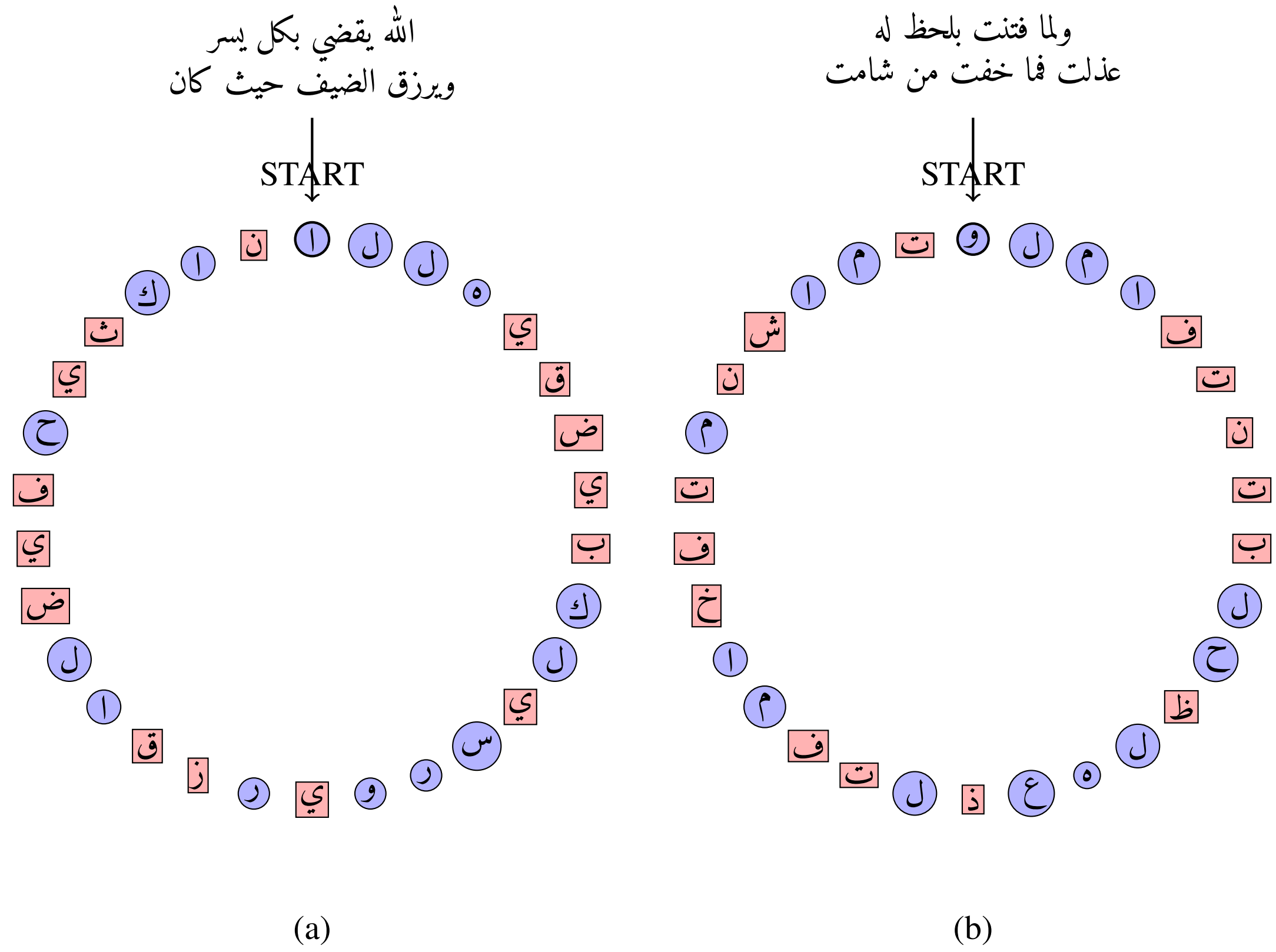

Figure 1: Two Arabic verses with 30 letters each, arranged clockwise on circles. Dotless letters are shown within circles, dotted letters within rectangles. In both cases, every ninth position eliminated (step size $k = 9$) is within rectangles, leaving only the 15 dotless letters. The verse in (b) was created by al-Ṣafadī.

*Associated Numerical Sequence:* We can associate with the sentence an integer sequence that records the consecutive runs of letters from each sub-alphabet. Beginning with the first letter, we count how many successive letters belong to the same sub-alphabet, then switch to the other sub-alphabet and continue counting. This way, the Arabic verse produces the sequence $(4, 5, 2, 1, 3, 1, 1, 2, 2, 3, 1, 2, 2, 1)$, reflecting the alternating structure of dotless and dotted letters as indicated in Figure 1. The sequence shows the number of letters from each sub-alphabet in each block of the sentence, where a block is a maximal string of letters from the same sub-alphabet.

[5] In standard Arabic orthography, *alif maqṣūrah* (ى) appears only in word-final position.

In the following figure, other verses by al-Ṣafadī from [5] are presented, each arranged on a circle.

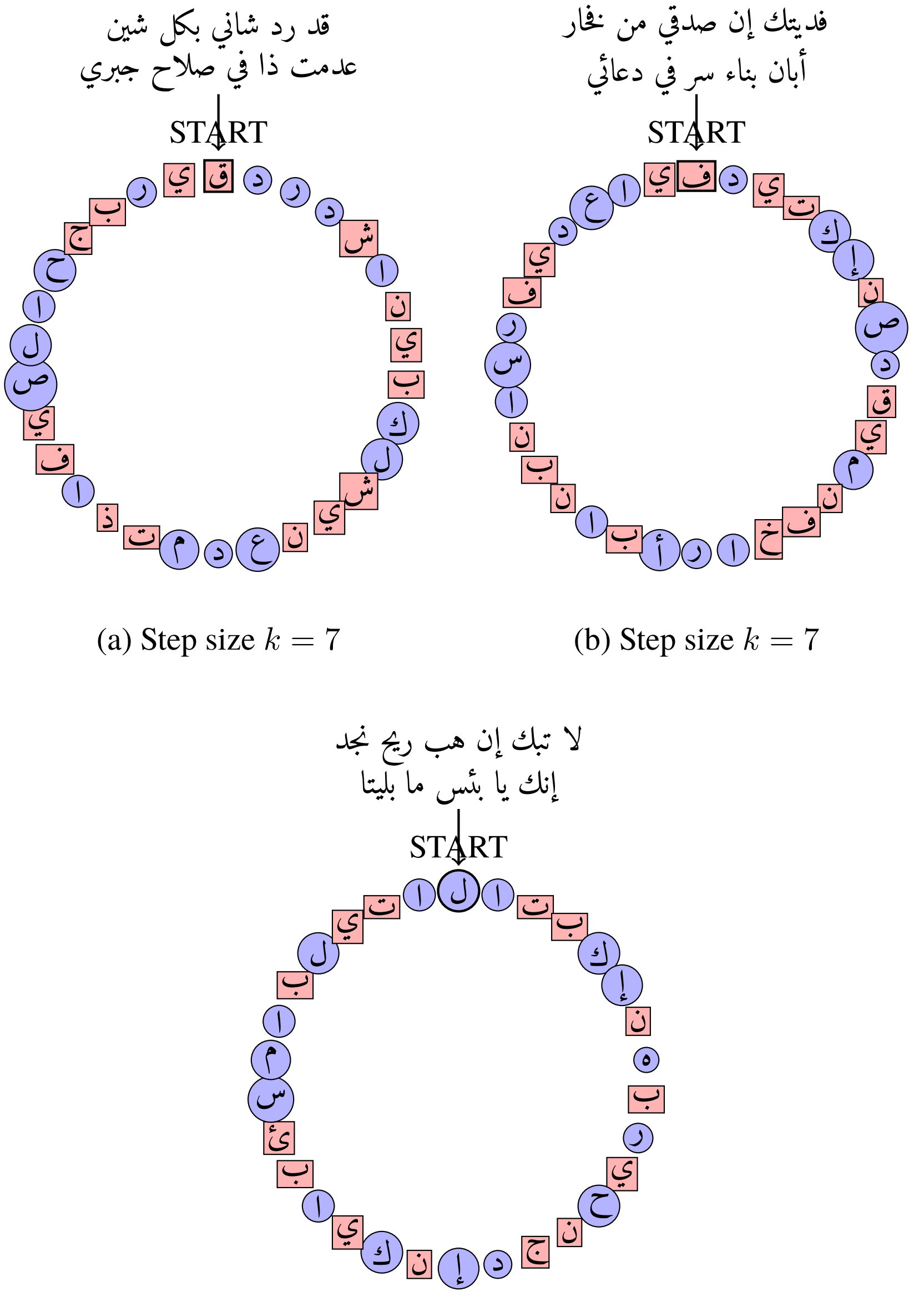

(a) Step size $k = 7$

(b) Step size $k = 7$

(c) Step size $k = 5$

Figure 2: Three Arabic verses arranged clockwise on circles. (a) 30 letters (15 dotless, 15 dotted) with step size $k = 7$. (b) 32 letters (16 dotless, 16 dotted) with step size $k = 7$. (c) 30 letters (16 dotless, 14 dotted) with step size $k = 5$. Dotless letters (including alif variations , ( are shown within circles, dotted letters within rectangles.

The integer sequence of the verse in Figure 2 (a) is $(1, 3, 1, 1, 3, 2, 3, 3, 2, 1, 2, 4, 2, 1, 1)$, the integer sequence of the verse in Figure 2 (b) is $(1, 1, 2, 2, 1, 2, 2, 1, 3, 3, 1, 1, 3, 3, 2, 3, 1)$ and the integer sequence of the verse in Figure 2 (c) is $(2, 2, 2, 1, 1, 1, 1, 1, 1, 2, 2, 1, 1, 1, 1, 2, 3, 1, 1, 2, 1)$, reflecting the alternating structure of their dotless and dotted letters.

### 3.1 Adaptation to the English Alphabet

The same idea can be adapted from Arabic to English by introducing an analogous partition of the English alphabet. Instead of dotless and dotted letters, we divide the capital letters of the English alphabet into two subgroups according to their geometric form: those composed purely of straight lines, and those containing at least one curved line. This yields the partition

$$\text{Straight-line group (15): } \{A, E, F, H, I, K, L, M, N, T, V, W, X, Y, Z\},$$

$$\text{Curved-line group (11): } \{B, C, D, G, J, O, P, Q, R, S, U\}.$$

To illustrate, suppose we have eight individuals to be divided into two groups of four. We use the sentence

"WE BURDEN",

which contains exactly eight letters. The individuals are arranged around a circle, each assigned to a letter of the sentence in order. Beginning from the first letter, we remove individuals on the circle using the step size $5$ (corresponding to removing every sixth person). We stop the elimination once four individuals remain. In this case, the survivors form two balanced groups, each containing two people. Figure 2 is an illustration of this example.

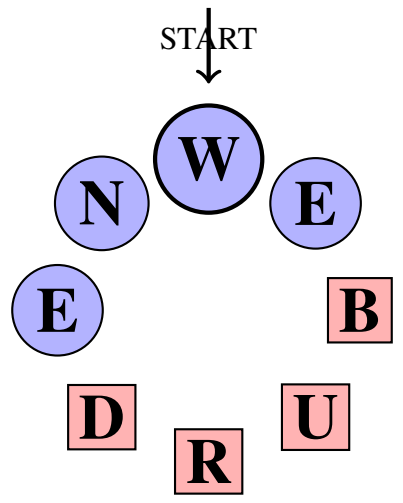

Figure 3: The 8 letters of the English sentence "WE BURDEN" placed compactly around a circle. Straight-line letters are shown within circles, curved-line letters within rectangles. With step size $5$, the 6th position is the first to be eliminated.

In this example, the associated sequence is $(2, 4, 2)$. In Section 5 (Figure 4), we provide more examples of English sentences with the associated numerical sequences $(4, 5, 2, 1, 3, 1, 1, 2, 2, 3, 1, 2, 2, 1)$, $(1, 3, 1, 1, 3, 2, 3, 3, 2, 1, 2, 4, 2, 1, 1)$, $(1, 1, 2, 2, 1, 2, 2, 1, 3, 3, 1, 1, 3, 3, 2, 3, 1)$, and $(2, 2, 2, 1, 1, 1, 1, 1, 1, 2, 2, 1, 1, 1, 1, 2, 3, 1, 1, 2, 1)$. Formalizing the problem in Section 5 and generalizing the Context-Free Grammar to formal languages will help us develop algorithms to determine whether sentences for a given sequence exist.

## 4 Generalized run length encoding

Motivated by the numerical sequence in the historical example, we introduce a generalization of the run-length encoding. We first define a partition of the alphabet.

[Partitioned Alphabet] Let $A$ be a finite alphabet. A collection $\{A_1, A_2, \ldots, A_r\}$ of non-empty subsets of $A$ is called a *partitioned alphabet* if:

1. The subsets are pairwise disjoint: $A_i \cap A_j = \emptyset$ for all $i \neq j$,
2. Each subset is non-empty: $A_i \neq \emptyset$ for all $i$.

We denote this as $(A; A_1, \ldots, A_r)$. If, in addition, $\bigcup_{i=1}^{r} A_i = A$, we say that the partitioned alphabet is *complete*, and $\{A_1, \ldots, A_r\}$ forms a partition of $A$.
Now, we state our generalization as follows.
[Generalized Run-Length Encoding] Let $(A; A_1, A_2, \ldots, A_r)$ be a partitioned alphabet. A *generalized run-length encoding* (GRL encoding) over this partitioned alphabet is a sequence

$$v = ((C_1, k_1), (C_2, k_2), \ldots, (C_m, k_m))$$

where:

1. Each $C_j$ is a multi-set of letters from some $A_{i_j}$ i.e., $\{C_j\} \subseteq A_{i_j}$ for some $i_j \in \{1, \ldots, r\}$, where $\{C_j\}$ denotes the underlying set of the multi-set $C_j$ (ignoring multiplicities)
2. Each $k_j = |C_j|$ is the cardinality of the multi-set $C_j$,
3. For each $j$, the index $i_j \in \{1, \ldots, r\}$ is uniquely determined by the condition $\{C_j\} \subseteq A_{i_j}$
4. $i_j \neq i_{j+1}$ for all $j = 1, 2, \ldots, m-1$ (consecutive runs come from different subsets of the partition).

The *length* of the GRL encoding is $n = k_1 + k_2 + \cdots + k_m$, and $m$ is called the *number of runs*.

**Remark 4.** *For a partitioned alphabet with $r = 2$ subsets $A_1$ and $A_2$, the alternating condition can be stated as:*

$$\textit{either } \{C_{2i-1}\} \subseteq A_1 \textit{ and } \{C_{2i}\} \subseteq A_2, \quad \textit{or} \quad \{C_{2i-1}\} \subseteq A_2 \textit{ and } \{C_{2i}\} \subseteq A_1.$$

*This explicitly shows the alternation between the two subsets.*

**Example 4.** *Consider the alphabet $A = \{a, b, c, d, e\}$ with partitioned alphabet $(A; A_1, A_2)$ where $A_1 = \{a, b\}$ and $A_2 = \{c, d, e\}$. For the word $w = aabccdd$, the GRL vector is $v = ((C_1, 3), (C_2, 4))$ where:*

- *$C_1 = \{a, a, b\}$ is a multi-set with $|C_1| = 3$ and $\{C_1\} = \{a, b\} \subseteq A_1$*
- *$C_2 = \{c, c, d, d\}$ is a multi-set with $|C_2| = 4$ and $\{C_2\} = \{c, d\} \subseteq A_2$*

For some applications, particularly in arrangement problems and combinatorial counting, we do not need to track the specific letters in each run, only which subset they come from and how many there are. This motivates the following abstraction:
[Partition-Level Generalized Run-Length Encoding] Let $(A; A_1, A_2, \ldots, A_r)$ be a partitioned alphabet. An *abstract generalized run-length vector* (abstract GRL vector) over this partitioned alphabet is a sequence

$$v = ((A_{i_1}, s_1), (A_{i_2}, s_2), \ldots, (A_{i_m}, s_m))$$

where:

1. Each $A_{i_j} \in \{A_1, A_2, \ldots, A_r\}$, $i_j$ indicates the subset,
2. Each $s_j \in \mathbb{Z}^+$ is a positive integer (the run length),
3. $i_j \neq i_{j+1}$ for all $j = 1, 2, \ldots, m-1$ (consecutive runs from different subsets).

The tuple $(A_{i_j}, s_j)$ represents $s_j$ consecutive positions labeled by the subset $A_{i_j}$, without specifying which particular letters from $A_{i_j}$ appear or their multiplicities.
The Partition-Level generalized run-length vector provides a natural encoding for a fundamental combinatorial problem: arranging $n$ distinguishable positions in a row, where each position is assigned to one of $r$ groups. An arrangement of length $n$ over $r$ groups is a sequence of $n$ positions, where each position is labeled with a group index from $\{1, 2, \ldots, r\}$. Each group index $i$ corresponds to subset $A_i$ in the partition. For example, with $n = 5$ positions and $r = 3$ groups, the arrangement $[1, 1, 2, 2, 3]$ assigns the first two positions to group 1 (subset $A_1$), the next two to group 2 (subset $A_2$), and the last position to group 3 (subset $A_3$).

The Partition-Level GRL vector naturally compresses such arrangements by encoding consecutive positions with the same subset label as a single run. The following theorem establishes that this encoding defines a bijection.

**Theorem 4.1.** *Let $n$ and $r$ be positive integers with $n \geq r$, and let $(A; A_1, \ldots, A_r)$ be a complete partitioned alphabet. The cardinality of various sets of partition-level GRL vectors of length $n$ is given as follows:*

- *General Case: The set of all partition-level GRL vectors of length $n$ over $(A; A_1, \ldots, A_r)$ has cardinality $r^n$.*
- *Surjective Case: The set of partition-level GRL vectors of length $n$ where all $r$ subsets appear at least once has cardinality*
$$\sum_{i=0}^{r-1} (-1)^i \binom{r}{i} (r-i)^n = r! \cdot S(n, r),$$
*where $S(n, r)$ is the Stirling number of the second kind.*
- *Balanced Case: Write $n = qr + s$ where $q = \lfloor n/r \rfloor$ and $0 \leq s < r$. The set of balanced partition-level GRL vectors of length $n$ (where exactly $s$ subsets appear $q+1$ times and the remaining $r - s$ subsets appear $q$ times) has cardinality*
$$\binom{r}{s} \cdot \frac{n!}{(q+1)!^s \cdot q!^{r-s}}.$$
*When $r$ divides $n$ (i.e., $s = 0$), this reduces to $\frac{n!}{(q!)^r}$ where $q = n/r$.*

*Proof.* Let's first establish a bijection with an appropriate set of arrangements. As defined above, an arrangement of length $n$ over $r$ groups is a sequence $a = (g_1, g_2, \ldots, g_n)$ where each $g_k \in \{1, 2, \ldots, r\}$ is a group index. Position $k$ is assigned to group $g_k$, which corresponds to subset $A_{g_k}$ of the partition. For convenience, we write arrangements using bracket notation: $a = [g_1, g_2, \ldots, g_n]$.

We define two maps between arrangements and partition-level GRL vectors as follows.

- *The map $\phi$*: Given an arrangement $a = [g_1, g_2, \ldots, g_n]$, we construct a partition-level GRL vector. Define positions $1 = p_1 < p_2 < \cdots < p_{m+1} = n+1$ where $p_{k+1}$ is the smallest position greater than $p_k$ such that $g_{p_{k+1}} \neq g_{p_k}$ (or $p_{k+1} = n+1$ if no such position exists). Then

$$\phi(a) = ((A_{g_{p_1}}, p_2 - p_1), (A_{g_{p_2}}, p_3 - p_2), \ldots, (A_{g_{p_m}}, p_{m+1} - p_m)).$$

  Note that $A_{g_{p_j}}$ denotes the subset corresponding to group index $g_{p_j}$, and $p_{j+1} - p_j$ is the length of the run.

- *The map $\psi$*: Given a partition-level GRL vector

$$v = ((A_{i_1}, s_1), (A_{i_2}, s_2), \ldots, (A_{i_m}, s_m)),$$

  we reconstruct an arrangement by expanding each run:

$$\psi(v) = [\underbrace{i_1, \ldots, i_1}_{s_1 \text{ times}}, \underbrace{i_2, \ldots, i_2}_{s_2 \text{ times}}, \ldots, \underbrace{i_m, \ldots, i_m}_{s_m \text{ times}}].$$

  Here, each $(A_{i_j}, s_j)$ in the GRL vector corresponds to a run of $s_j$ consecutive positions labeled with group index $i_j \in \{1, 2, \cdots, r\}$.

We verify that $\phi$ and $\psi$ are bijections and inverse maps by showing $\psi \circ \phi = \text{id}$ and $\phi \circ \psi = \text{id}$, respectively. Let $a = [g_1, \ldots, g_n]$ be an arrangement. The $\phi(a)$ identifies all maximal runs. The map $\psi$ reconstructs exactly these runs, recovering the original sequence $a$. Thus, we have $\psi \circ \phi = \text{id}$. Now, let $v = ((A_{i_1}, s_1), \ldots, (A_{i_m}, s_m))$ be a partition-level GRL vector. The map $\psi$ creates an arrangement with $s_1$ consecutive positions labeled $i_1$, followed by $s_2$ consecutive positions labeled $i_2$, and so on. Compressing this arrangement back using $\phi$, we identify exactly these same $m$ runs, recovering the original vector $v$. Hence, $\phi \circ \psi = \text{id}$.

Therefore, $\phi$ and $\psi$ establish a bijection between arrangements and partition-level GRL vectors. Now we prove each part by counting the corresponding set of arrangements.

Proof of Part (i): The set of all arrangements of length $n$ over $r$ groups has cardinality $r^n$ since each of the $n$ positions independently chooses one of $r$ group indices from $\{1, 2, \ldots, r\}$. Hence, by the bijection $\phi$, the set of all partition-level GRL vectors of length $n$ has the same cardinality $r^n$.

Proof of Part (ii): Let $\mathcal{A}_{\text{surj}}$ denote the set of arrangements where all $r$ groups appear at least once, and let $\mathcal{G}_{\text{surj}}$ denote the set of partition-level GRL vectors where all $r$ subsets appear. We show that the bijection $\phi$ restricts to a bijection $\mathcal{A}_{\text{surj}} \to \mathcal{G}_{\text{surj}}$. If an arrangement $a$ uses all $r$ group indices, then $\phi(a)$ contains at least one run from each subset. And conversely, if a GRL vector $v$ contains runs from all $r$ subsets, then $\psi(v)$ uses all $r$ group indices.

We count $|\mathcal{A}_{\text{surj}}|$ using inclusion-exclusion. For $I \subseteq \{1, 2, \ldots, r\}$, let $B_I$ be the set of arrangements that do not use any group index $i$ for $i \in I$. Then $|B_I| = (r - |I|)^n$ since we can only use the $r - |I|$ indices not in $I$.

By inclusion-exclusion:

$$\begin{aligned}|\mathcal{A}_{\text{surj}}| &= r^n - \sum_{\emptyset \neq I \subseteq \{1,\dots,r\}} (-1)^{|I|+1} |B_I| \\ &= r^n + \sum_{\emptyset \neq I \subseteq \{1,\dots,r\}} (-1)^{|I|} (r - |I|)^n \\ &= \sum_{i=0}^{r} (-1)^i \binom{r}{i} (r-i)^n \\ &= \sum_{i=0}^{r-1} (-1)^i \binom{r}{i} (r-i)^n,\end{aligned}$$

where the last equality uses $(r-r)^n = 0$ for $n \geq 1$. This expression is equal to $r! \cdot S(n,r)$ by the well-known identity relating Stirling numbers of the second kind to the number of surjective functions.

Proof of Part (iii): Let $\mathcal{A}_{\text{bal}}$ denote the set of arrangements where exactly $s$ group indices appear $q+1$ times and the remaining $r-s$ indices appear $q$ times, and let $\mathcal{G}_{\text{bal}}$ denote the set of balanced partition-level GRL vectors with the same count distribution. The bijection $\phi$ restricts to a bijection $\mathcal{A}_{\text{bal}} \to \mathcal{G}_{\text{bal}}$ since the $\phi$ and $\psi$ maps preserve the total count for each group index. If arrangement $a$ has count vector $(k_1, \dots, k_r)$ (where $k_i$ is the total number of positions labeled $i$), then $\phi(a)$ has the same count vector, since mapping consecutive runs does not change the total count for each index. Similarly, $\psi$ preserves the count vectors.

To count $|\mathcal{A}_{\text{bal}}|$, we first choose $s$ group indices out of $\{1, \dots, r\}$ that appear $q+1$ times. We can do it in $\binom{r}{s}$ ways. Then, for a fixed choice, say indices $1, \dots, s$ appear $q+1$ times and indices $s+1, \dots, r$ appear $q$ times, we count arrangements of $n$ positions with this distribution. Then, the number of such arrangements is the multinomial coefficient

$$\frac{n!}{(q+1)!^s \cdot q!^{r-s}}.$$

Multiplying these gives

$$|\mathcal{A}_{\text{bal}}| = \binom{r}{s} \cdot \frac{n!}{(q+1)!^s \cdot q!^{r-s}}.$$

When $s = 0$, we have $(q+1)!^0 = 1$ and all indices appear exactly $q = n/r$ times, giving cardinality $\dfrac{n!}{(q!)^r}$. □

**Remark 5.** *The bijections in Theorem 4.1 provide a combinatorial interpretation that each partition-level GRL vector corresponds to a unique way of assigning $n$ positions to $r$ groups, where group $i$ corresponds to subset $A_i$ of the partition.*

# 5 From Examples to Formalization

In the preceding sections, we illustrated some partitioning procedures using concrete examples from the Arabic and English alphabets. We now turn to a formal description of the underlying problem.

Our aim is to abstract the essential ingredients –an alphabet, a partitioning rule, sequences encoding alternation between sub-alphabets, and a lexicon with grammatical structure– into a unified framework. This enables us to pose precise questions about the existence, construction, and enumeration of valid sentences that adhere to both linguistic and combinatorial constraints. Formally, the problem setup consists of the following components.

A *Run-Constrained Sentence Instance* consists of:

- An *alphabet* $\Sigma$ with a *partition function* $\Pi : \Sigma \to \{A, B\}$ dividing symbols into two classes $A = \Pi^{-1}(A)$ and $B = \Pi^{-1}(B)$.
- A *length pattern* $(k_1, k_2, \ldots, k_m)$ specifying the required block sizes.
- A finite *dictionary* $D \subseteq \Sigma^*$ of meaningful words, where $\Sigma^*$ denotes the set of all finite strings over alphabet $\Sigma$, including the empty string $\varepsilon$.
- A *grammar* $G$ over $D$ specifying which word sequences are grammatical.

Given this setup, a fundamental question for our purposes can be formulated as follows.

**Problem 1** (Main Problem: Find a Valid Sequence and Sentence)**.** *Given a Run-Constrained Sentence Instance* $(\Sigma, \Pi, (k_1, \ldots, k_m), D, G)$*, find:*

1. *An alternating sequence* $t = ((C_1, k_1), (C_2, k_2), \ldots, (C_m, k_m))$ *where each* $C_i$ *is a multi-set with* $|C_i| = k_i$*, and either*

$$\{C_{2i}\} \subseteq A \textit{ and } \{C_{2i-1}\} \subseteq B, \quad \textit{or} \quad \{C_{2i-1}\} \subseteq A \textit{ and } \{C_{2i}\} \subseteq B,$$

   *where* $\{C_i\}$ *denote the set of distinct symbols in multi-set* $C_i$ *for* $1 \leq i \leq m$.

2. *Let* $D^*$ *be the set of all finite sequences of words from the dictionary* $D$*. A sentence* $w \in D^*$ *such that:*
   - $w$ *is grammatical with respect to grammar* $G$*, and*
   - *The concatenated letters of* $w$ *(obtained by spelling out each word) satisfy the block pattern defined by* $t$*.*

*If no such pair* $(t, w)$ *exists, return null.*

We make two observations about the structure of this problem.

**Remark 6.** *Problem 1 can be viewed as having two components:*

1. *Sequence construction: Find an alternating sequence* $t$ *that satisfies the alternation constraint.*
2. *Sentence generation: For a given alternating sequence* $t$*, find a grammatical sentence* $w$ *whose letter pattern matches* $L_t$*.*

**Remark 7.** *In the following subsection, we formalize the conditions in Problem 1 part 2, using formal language theory: condition (1) becomes* $w \in L(G)$ *where* $L(G)$ *denotes the language generated by grammar* $G$*, and condition (2) becomes* $h(w) \in L_t$ *where* $h : D^* \to \Sigma^*$ *is the spelling homomorphism mapping word sequences to letter sequences, and* $L_t$ *is the block language defined by sequence* $t$*.*

## 5.1 Formal Language Theory Background

In this section, we review some fundamental concepts of formal language theory. The following definitions from [2] are standard in automata theory.

[ [2], Deterministic Finite Automaton] A *deterministic finite automaton (DFA)* is a 5-tuple $M = (Q, \Sigma, \delta, q_0, F)$ where:

- $Q$ is a finite set of states,
- $\Sigma$ is a finite alphabet,
- $\delta : Q \times \Sigma \to Q$ is the transition function,
- $q_0 \in Q$ is the start state,
- $F \subseteq Q$ is the set of accepting (final) states.

A DFA $M$ accepts a string $w = w_1 w_2 \cdots w_n \in \Sigma^*$ if the sequence of states $q_0, q_1, \ldots, q_n$ defined by $q_{i+1} = \delta(q_i, w_{i+1})$ satisfies $q_n \in F$.

We define the language recognized by a DFA as follows.

[ [2], Regular Language] Let $M = (Q, \Sigma, \delta, q_0, F)$ be a deterministic finite automaton (DFA). The *language recognized by* $M$ is

$$L(M) = \{\, w \in \Sigma^* \mid M \text{ accepts } w \,\},$$

where $\delta(q_0, w)$ denotes the state reached by $M$ after processing the string $w$ from the start state $q_0$. A language $L \subseteq \Sigma^*$ is called *regular* if there exists a DFA $M$ such that $L = L(M)$.

Having established the basics of finite automata, we now turn to more expressive grammatical formalisms. The following definition generalizes the standard notion of context-free grammars to operate over dictionaries of words rather than individual symbols.

[ [2], Context-Free Grammar over a Dictionary] A *context-free grammar* $G = (V, D, P, S)$ over a dictionary $D \subseteq \Sigma^*$ consists of:

- $V$ is a finite set of nonterminal symbols,
- $D$ is the dictionary (terminals are words, each word $w \in D$ is a string over $\Sigma$),
- $P$ is a finite set of production rules $A \to \alpha$ where $A \in V$ and $\alpha \in (V \cup D)^*$,
- $S \in V$ is the start symbol.

The language $L(G) \subseteq D^*$ is the set of word sequences derivable from $S$.

Now, we recall the definition of the class of languages generated by context-free grammars.

[ [2], Context-Free Language] A language $L$ is called a *context-free language (CFL)* if there exists a context-free grammar $G$ such that $L = L(G)$ where $L(G)$ is the language of $G$.

## 5.2 Block Language and Its Properties

Based on our formulation of the main problem, we define a block language as follows.
[Block Language] For a sequence $t = ((C_1, k_1), \ldots, (C_m, k_m))$, let $\Sigma_{C_i} = \{C_i\}$ denote the set of distinct symbols in multi-set $C_i$. The *block language* is defined as

$$L_t = (\Sigma_{C_1})^{k_1}(\Sigma_{C_2})^{k_2} \cdots (\Sigma_{C_m})^{k_m} \subseteq \Sigma^K, \quad K = \sum_{i=1}^{m} k_i.$$

where the language $(\Sigma_{C_i})^{k_i}$ consists of all strings of length $k_i$ using symbols from $\Sigma_{C_i}$, which is a finite set containing at most $|\Sigma_{C_i}|^{k_i}$ strings.
Having defined block languages, we now establish their key computational property.

**Theorem 5.1.** *For every sequence* $t = \big((C_1, k_1), (C_2, k_2), \ldots, (C_m, k_m)\big)$ *as defined above, the language* $L_t$ *is regular.*

*Proof.* For each $i \in \{1, 2, \ldots, m\}$, the set $\Sigma_{C_i}$ is a finite subset of the alphabet, and by Definition 5.1, every finite language is regular. The block language $L_t = (\Sigma_{C_1})^{k_1}(\Sigma_{C_2})^{k_2} \cdots (\Sigma_{C_m})^{k_m}$ is a concatenation of $m$ regular languages. Since the class of regular languages is closed under finite concatenation, $L_t$ is regular. □

Since every regular language is context-free, we immediately obtain the following result.

**Corollary 5.2.** *For every sequence* $t = \big((C_1, k_1), (C_2, k_2), \ldots, (C_m, k_m)\big)$ *as defined above,* $L_t$ *is a context-free language (CFL).*

*Proof.* Since every regular language is a context-free language by [2, Exercise 5.1.3], $L_t$ is a context-free language. □

Now, let's investigate the decidability of our main problem in this section.

Our problem operates on two levels:

1. Letter level: Each word $w \in D$ is a string over the alphabet $\Sigma$. The block constraint $L_t$ operates at this level, prescribing an alternation pattern for the concatenated letters.
2. Word level: The grammar $G$ operates on words as atomic units, generating sequences in $D^*$.

Suppose $h : D^* \to \Sigma^*$ is a homomorphism (with respect to the concatenation operation) that bridges these levels by mapping each word sequence to its concatenated letter sequence. Then, our problem asks:

Does there exist a word sequence $w \in D^*$ such that

- $w \in L(G)$ (grammatical word sequence), and
- $h(w) \in L_t$ (concatenated letters satisfy the alternation pattern)?

Equivalently, is $L(G) \cap h^{-1}(L_t) \neq \emptyset$, or is $h(L(G)) \cap L_t \neq \emptyset$?

**Lemma 5.3** (Homomorphic image of CFLs)**.** *Let $L(G) \subseteq D^*$ be a context-free language and let $h : D^* \to \Sigma^*$ be the homomorphism that maps each word to its spelling (concatenation of its letters). Then $h(L(G))$ is context-free.*

*Proof.* Context-free languages are invariant under homomorphism [2, Theorem 7.24]. Since $h$ is a string homomorphism and $L(G)$ is context-free, $h(L(G))$ is context-free. □

We now combine the structural properties of block languages with the homomorphism lemma to establish our main decidability result.

**Theorem 5.4** (Decidability of Existence)**.** *The existence problem (Problem 1) is decidable in polynomial time.*

*Proof.* By Theorem 5.1, $L_t$ is a regular language. By Lemma 5.3, $h(L(G))$ is a context-free language, where $h : D^* \to \Sigma^*$ maps each word to its spelling. The existence problem asks whether $h(L(G)) \cap L_t \neq \emptyset$. Since context-free languages are closed under intersection with regular languages [2, Theorem 7.27], $h(L(G)) \cap L_t$ is a context-free language. The emptiness problem for context-free languages is decidable in linear time in the size of the grammar [2, Section 7.4.3]. Therefore, the existence problem is decidable in polynomial time. □

### 5.3 Word-Level Automaton Construction

To establish the regularity of block languages and enable our decidability result, we construct explicit automata that recognize sequences satisfying block constraints. We begin with a letter-level DFA operating on the alphabet $\Sigma$.

[DFA for Block Language] For a sequence $t = ((C_1, k_1), \ldots, (C_m, k_m))$, we construct a DFA $M_t = (Q, \Sigma, \delta, q_0, F)$ that recognizes $L_t$ as follows:

- $Q = \{q_0, q_1, \ldots, q_K\}$ with $K = \sum_{i=1}^{m} k_i$ (states represent positions 0 through $K$),
- Start state is $q_0$ (no letters read yet),
- Accepting states $F = \{q_K\}$ (exactly $K$ letters read),
- Transition function $\delta$ defined as: For state $q_j$ where $0 \leq j < K$ and symbol $\sigma \in \Sigma$:

  $$\delta(q_j, \sigma) = \begin{cases} q_{j+1} & \text{if } \sigma \in \Sigma_{C_i} \text{ where } i \text{ is such that } \sum_{\ell=1}^{i-1} k_\ell < j+1 \leq \sum_{\ell=1}^{i} k_\ell \\ \text{reject} & \text{otherwise} \end{cases}$$

  In other words, from position $j$, we can read symbol $\sigma$ and advance to position $j+1$ if and only if position $j+1$ falls within a block $i$ for which $\sigma \in \Sigma_{C_i}$.

- For the accepting state $q_K$, we define $\delta(q_K, \sigma) =$ reject for all $\sigma \in \Sigma$.

While $M_t$ operates on individual letters, our main problem involves a context-free grammar over a dictionary $D$ of words. To bridge this gap, we lift the letter-level DFA to a word-level automaton that tracks which dictionary words are compatible with the block constraints.

[Word-Level automaton] Given a DFA $M_t = (Q, \Sigma, \delta_t, q_0, F)$ recognizing block language $L_t$ and a dictionary $D \subseteq \Sigma^*$, a *word-level automaton* is a directed graph $\mathcal{W} = (Q, D, E)$ where:

- $Q$ is the state set from $M_t$ (representing positions in the letter sequence),
- $D$ is the dictionary (edge labels),
- $E \subseteq Q \times D \times Q$ is the edge set defined by:

  $$(q, w, q') \in E \quad \text{if and only if} \quad \delta_t^*(q, w) = q'$$

  where $\delta_t^* : Q \times \Sigma^* \to Q$ is the extended transition function that reads strings (i.e., reading the letters of the word $w$ from state $q$ in $M_t$ leads to state $q'$).

The word-level automaton $\mathcal{W}$ has the same states as the letter-level DFA $M_t$, but its transitions are labeled with entire words rather than single letters. An edge $(q, w, q') \in E$ exists if and only if:

- Starting from state $q$ in $M_t$,
- Reading the letters of word $w = w_1 w_2 \cdots w_\ell$ one by one,
- We successfully reach state $q'$ in $M_t$.

**Remark 8.** *The word-level automaton $\mathcal{W}$ effectively "lifts" the letter-level DFA $M_t$ to operate on words as atomic units. A path from $q_0$ to an accepting state in $\mathcal{W}$ corresponds to a sequence of words whose concatenated letters form a string in $L_t$.*

### 5.4 Algorithm for Finding Valid Sentences

We now present the algorithmic realization of our decidability result, which proceeds in two stages: first, checking existence for a given sequence, then searching over all possible sequences.

**Algorithm 1** Check Existence for a Given Sequence

**Require:** (Input) Sequence $t = ((C_1, k_1), \ldots, (C_m, k_m))$, dictionary $D$, grammar $G$
**Ensure:** (Output) A valid sentence $w$ if one exists, **null** otherwise

Construct DFA $M_t = (Q, \Sigma, \delta_t, q_0, F)$ recognizing $L_t$ with $K + 1$ states where $K = \sum_{i=1}^{m} k_i$
Initialize word-level automaton $\mathcal{W} = (Q, D, \emptyset)$ with empty edge set
**for** each word $w \in D$ **do**
    **for** each state $q \in Q$ **do**
        Simulate reading letters of $w$ from state $q$ in $M_t$
        **if** simulation reaches state $q'$ without rejection **then**
            Add edge $(q, w, q')$ to $E$ in $\mathcal{W}$
        **end if**
    **end for**
**end for**
Apply CFG parsing algorithm (e.g., CKY [7]) on $\mathcal{W}$ to find a grammatical path from $q_0$ to accepting state in $F$
**if** grammatical path $p$ exists **then**
    Extract sentence $w$ from path $p$ (sequence of word labels on edges)
    **return** $w$
**else**
    **return null**
**end if**

The following theorem establishes the complexity of Algorithm 1.

**Theorem 5.5.** *Algorithm 1 runs in time $O(|D| \cdot |t| + |t|^3 \cdot |G|)$, where $|t| = \sum_{i=1}^{m} k_i$ is the total pattern length, assuming constant alphabet size $|\Sigma|$.*

*Proof.* Let $K = \sum_{i=1}^{m} k_i$ denote the total pattern length. We analyze the runtime of each component of the algorithm. Constructing the DFA $M_t$ requires creating $K + 1$ states with transitions defined according to Definition 5.3. For each of the $K$ non-accepting states and each symbol in $\Sigma$, we must determine the appropriate transition, which takes $O(K \cdot |\Sigma|)$ time. Building the word-level automaton involves testing each word $w \in D$ from each state $q \in Q$. For a word $w$ of length $\ell_w$, simulating its reading from a given state requires $O(\ell_w)$ time. The total time across all words and all states is

$$O\left(\sum_{w\in D} |Q| \cdot \ell_w\right) = O(K \cdot \sum_{w\in D} \ell_w) = O(K \cdot |D|),$$

where $|D| = \sum_{w\in D} \ell_w$ denotes the total number of letters across all words in the dictionary. Applying the CKY parsing algorithm on the word-level automaton with $K + 1$ positions requires $O(K^3 \cdot |G|)$ time, where $|G|$ is the size of the grammar. If a valid sentence exists, extracting it from the parse structure takes at most $O(K)$ additional time. Summing the runtimes and treating $|\Sigma|$ as constant, we obtain $O(K \cdot |\Sigma| + K \cdot |D| + K^3 \cdot |G|) = O(|D| \cdot |t| + |t|^3 \cdot |G|)$, where we take $K = O(|t|)$. □

To solve Problem 1 completely, we must find both a valid sequence $t$ and a corresponding sentence. The following algorithm combines sequence enumeration with the checking procedure.

**Algorithm 2** Find a Valid Alternating Sequence and a Sentence

**Require:** Alphabet partition $(A, B)$; length pattern $(k_1, \ldots, k_m)$; dictionary $D$; grammar $G$
**Ensure:** A pair $(t, w)$ where $t$ is a valid sequence and $w$ is a valid sentence, or **null** if none exists

**for** $\alpha \in \{(A, B, A, \ldots), (B, A, B, \ldots)\}$ **do** ▷ Try both alternation patterns
    **for** each valid choice of multi-sets $(C_1, \ldots, C_m)$ **do**
        where $C_i$ has $|C_i| = k_i$ and $\{C_i\} \subseteq \alpha_i$
        Construct $t \leftarrow ((C_1, k_1), \ldots, (C_m, k_m))$
        $w \leftarrow$ Algorithm $1(t, D, G)$
        **if** $w \neq$ **null then**
            **return** $(t, w)$ ▷ Found valid sequence and sentence
        **end if**
    **end for**
**end for**
**return null** ▷ No valid sequence exists

The following theorem establishes the complexity of Algorithm 2.

**Theorem 5.6.** *Let $|A| = a$ and $|B| = b$ denote the sizes of the partition classes. Algorithm 2 runs in time*

$$O\left(N \cdot (|D| \cdot |t| + |t|^3 \cdot |G|)\right)$$

*where $N$ is the number of possible sequences $t$, $|\alpha_i| \in \{a, b\}$, and $N \leq 2 \cdot \prod_{i=1}^{m} \binom{|\alpha_i| + k_i - 1}{k_i}$, which can be exponential in $k_i$ and $|\Sigma|$.*

*Proof.* We first determine the number of sequences $t$ that Algorithm 2 must check, then analyze the cost per sequence.
The algorithm considers two alternation patterns: sequences starting with partition class $A$ alternating with $B$, and sequences starting with $B$ alternating with $A$. For each alternation pattern, we must enumerate all possible choices of multi-sets $(C_1, \ldots, C_m)$ satisfying the constraints. For a given alternation pattern, each block $i$ must use symbols from its designated partition class $\alpha_i \in \{A, B\}$, where $|\alpha_i| \in \{a, b\}$. Block $i$ requires a multi-set $C_i$ of size $k_i$ drawn from an alphabet of size $|\alpha_i|$. The number of such multi-sets equals the number of ways to choose $k_i$ items with repetition from $|\alpha_i|$ distinct symbols. Equivalently, it is the number of non-negative integer solutions to the equation $x_1 + x_2 + \cdots + x_{k_i} = |\alpha_i|$. It is well known that this number is given by the binomial coefficient $\left(\!\binom{|\alpha_i|}{k_i}\!\right) = \binom{|\alpha_i| + k_i - 1}{k_i}$ (one way of obtaining this identity is the "stars-and-bars" combinatorial argument). Since the choice of the multi-set for each block is independent, the total number of multi-set sequences for a fixed alternation pattern is the product $\prod_{i=1}^{m} \binom{|\alpha_i| + k_i - 1}{k_i}$. Accounting for both alternation patterns, the total number of sequences to check is $N = 2 \cdot \prod_{i=1}^{m} \binom{|\alpha_i| + k_i - 1}{k_i}$.
For each sequence $t$ constructed in this manner, the algorithm invokes Algorithm 1 to test whether a valid grammatical sentence exists and, if so, to construct it. By Theorem 5.5, each such invocation requires $O(|D| \cdot |t| + |t|^3 \cdot |G|)$ time. In the worst case, the algorithm must check all $N$ sequences before finding a valid one (or determining none exists). Therefore, the total runtime is $O(N \cdot (|D| \cdot |t| + |t|^3 \cdot |G|))$. Note that $N$ grows exponentially with the block sizes $k_i$ and the partition class sizes $a$ and $b$. Specifically, when $k_i$ is large or when $|\alpha_i|$ is large, the binomial coefficients $\binom{|\alpha_i|+k_i-1}{k_i}$ become exponentially large, making $N$ exponential in the worst case. □

We highlight several observations that can be used to optimize the algorithm.

**Remark 9.** *Several optimization ideas can reduce the search space:*

1. *Dictionary-guided pruning: Only consider multi-sets $C_i$ whose symbols actually appear in words from $D$.*
2. *Early termination: Stop as soon as any valid sequence is found.*
3. *Incremental construction: Build the lattice incrementally and prune sequences that cannot lead to valid sentences.*

We now formally establish the correctness of Algorithm 2 for Problem 1.

**Theorem 5.7.** *Algorithm 2 correctly solves Problem 1. If a valid pair $(t, w)$ exists, the algorithm returns such a pair. If no valid pair exists, the algorithm returns null.*

*Proof.* Algorithm 2 exhaustively searches all possible alternating sequences $t$ by enumerating both alternation patterns and all valid multi-set choices. For each candidate sequence $t$, it invokes Algorithm 1 to determine whether a grammatical sentence exists for that sequence. By Theorem 5.1, each $L_t$ is regular. By Lemma 5.3 and the proof of Theorem 5.4, Algorithm 1 correctly determines whether $h(L(G)) \cap L_t \neq \emptyset$ and, if so, returns a witness sentence $w \in h(L(G)) \cap L_t$. Since Algorithm 2 checks all possible sequences, it will find a valid pair $(t, w)$ if one exists. If the algorithm returns null, it means no alternating sequence admits a valid grammatical sentence, so no solution to Problem 1 exists. □

To verify our algorithm, we implemented it in Python and applied it to various test cases. The complete implementation is available in our GitHub repository[6]. One such example is the sentence "THAT SOURCE: ADVANCED CLASS RESULTS.", which was generated by the algorithm with the sequence $(4, 5, 2, 1, 3, 1, 1, 2, 2, 3, 1, 2, 2, 1)$ which is the sequence of the historical example. The algorithms also produce the sentence "HOUSE SITEPRINT: PROVIDE ACCOUNT... OH!" for the sequence $(1, 3, 1, 1, 3, 2, 3, 3, 2, 1, 2, 4, 2, 1, 1)$, the sentence "STORE, AUTHOR EDUCATION; POST THOSE INS." for the sequence $(1, 1, 2, 2, 1, 2, 2, 1, 3, 3, 1, 1, 3, 3, 2, 3, 1)$, and the sentence "DO MESSAGE RIGHT? COMPARE ISSUES - FAQ" for the sequence $(2, 2, 2, 1, 1, 1, 1, 1, 1, 2, 2, 1, 1, 1, 1, 2, 3, 1, 1, 2, 1)$.

Figure 4 illustrates how these letters of these sentences are arranged around a circle, with straight-line letters colored in blue and curved-line letters in red.

[6] github.com/omidkhormali/Historic_english_sentence_code

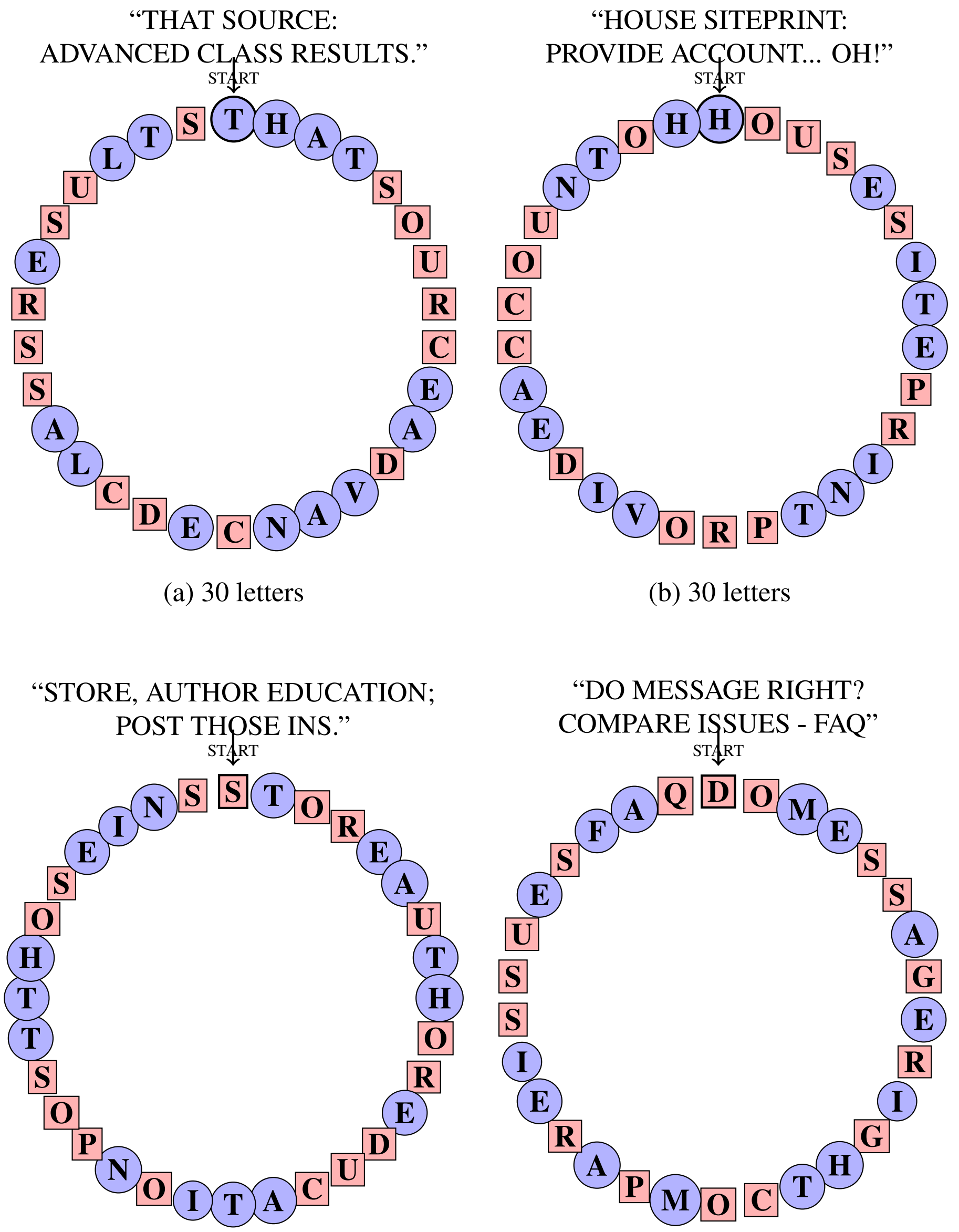

(a) 30 letters (b) 30 letters (c) 32 letters (d) 30 letters

Figure 4: Four English sentences arranged around circles. (a) "THAT SOURCE: ADVANCED CLASS RESULTS." with 30 letters following the historical example sequence. (b) "HOUSE SITEPRINT: PROVIDE ACCOUNT... OH!" with 30 letters. (c) "STORE, AUTHOR EDUCATION; POST THOSE INS." with 32 letters. (d) "DO MESSAGE RIGHT? COMPARE ISSUES - FAQ" with 30 letters. Straight-line letters are shown within circles; curved-line letters within rectangles.

These examples demonstrate that our algorithms can generate sentences in a different language using patterns from well-known historical examples.

**Remark 10.** *Python programs were run to implement the algorithms using a dictionary of the 2,000 most common words in English. For the historical sequence (30 letters, Figure 4 (a)), the algorithm identified a complete sentence that satisfied all constraints. For the other three sequences (30, 32, and 30 letters, in Figure 4 (b), (c,) and (d), respectively), the algorithm generated sentences of*

*28, 29, and 27 letters, respectively, which were then completed by manually adding short connecting words (2-3 letters each). Additionally, punctuation marks (commas, colons, and ellipses) were added manually to enhance readability and to create more coherent sentences. Note that the sequences depend only on letters and not on punctuation marks, so these additions do not affect the validity of the solutions. These results demonstrate the algorithm's effectiveness even with a limited dictionary. A more comprehensive lexicon would likely yield better solutions for all sequences without requiring manual completion.*

While the previous example demonstrates the algorithm's output on a historical pattern, we now provide a complete walkthrough of Algorithm 1 on a simpler instance to illustrate the computational steps in detail.

**Example 5.** *To illustrate the algorithm's operation in detail, we trace through a complete execution on a small instance.*

*Input specification.* *We consider the alphabet $\Sigma = \{a, b, c, d\}$ partitioned into $A = \{a, b\}$ and $B = \{c, d\}$. The block sequence is $t = ((C_1, 2), (C_2, 1), (C_3, 2))$ where $C_1 = \{a, a\}$ with $\Sigma_{C_1} = \{a\}$, $C_2 = \{c\}$ with $\Sigma_{C_2} = \{c\}$, and $C_3 = \{a, b\}$ with $\Sigma_{C_3} = \{a, b\}$. This yields a total pattern length $K = 5$. The block language is thus*

$$L_t = \{a\}^2 \cdot \{c\}^1 \cdot \{a, b\}^2 = \{aacaa, aacab, aacba, aacbb\}.$$

*The dictionary is $D = \{$"aa", "ac", "ab", "c", "ca"$\}$, and the grammar $G$ has start symbol $S$ with production rules:*

$$S \to W\,W\,W, \quad W \to \text{"aa"} \mid \text{"ac"} \mid \text{"ab"} \mid \text{"c"} \mid \text{"ca"}$$

*This grammar accepts any sequence of exactly three words from $D$.*

*Constructing the letter-level DFA.* *Following Definition 5.3, we construct $M_t = (Q, \Sigma, \delta_t, q_0, F)$ with states $Q = \{q_0, q_1, q_2, q_3, q_4, q_5\}$ representing positions $0$ through $5$, start state $q_0$, and accepting state $F = \{q_5\}$. The transition function $\delta_t$ is defined according to the block constraints:*

| ***State*** | ***Block*** | $\delta_t(\cdot, a)$ | $\delta_t(\cdot, b)$ | $\delta_t(\cdot, c)$ |
|---|---|---|---|---|
| $q_0$ | *Block 1 (next pos)* | $q_1$ | *reject* | *reject* |
| $q_1$ | *Block 1* | $q_2$ | *reject* | *reject* |
| $q_2$ | *Block 2 (next pos)* | *reject* | *reject* | $q_3$ |
| $q_3$ | *Block 3 (next pos)* | $q_4$ | $q_4$ | *reject* |
| $q_4$ | *Block 3* | $q_5$ | $q_5$ | *reject* |
| $q_5$ | *(accepting)* | *reject* | *reject* | *reject* |

*The DFA diagram is as follows.*

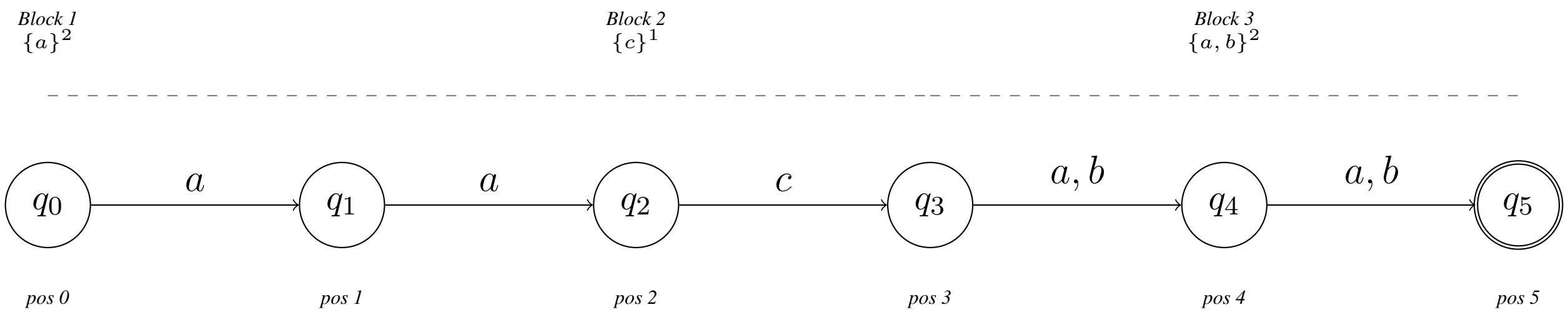

*From state $q_i$ (position $i$), we can read symbol $\sigma$ and transition to $q_{i+1}$ if and only if position $i+1$ falls within a block whose constraint set contains $\sigma$. For instance, from $q_0$, only symbol $a$ is valid since position 1 belongs to Block 1 with constraint set $\{a\}$. Similarly, from $q_3$, both $a$ and $b$ are valid since position 4 belongs to Block 3 with constraint set $\{a, b\}$.*

*Constructing the word-level automaton. For each word $w \in D$, we determine the edges $(q, w, q')$ in the word-level automaton $\mathcal{W}$ by simulating $M_t$ reading the letters of $w$ from each state $q$. We summarize the results:*

| ***Word*** | ***From State*** | ***To State*** | ***Letter Sequence*** |
|---|---|---|---|
| *"aa"* | $q_0$ | $q_2$ | $a, a$ |
| *"aa"* | $q_3$ | $q_5$ | $a, a$ |
| *"ac"* | $q_1$ | $q_3$ | $a, c$ |
| *"ab"* | $q_3$ | $q_5$ | $a, b$ |
| *"c"* | $q_2$ | $q_3$ | $c$ |
| *"ca"* | $q_2$ | $q_4$ | $c, a$ |

*The word-level atomaton diagram is as follows.*

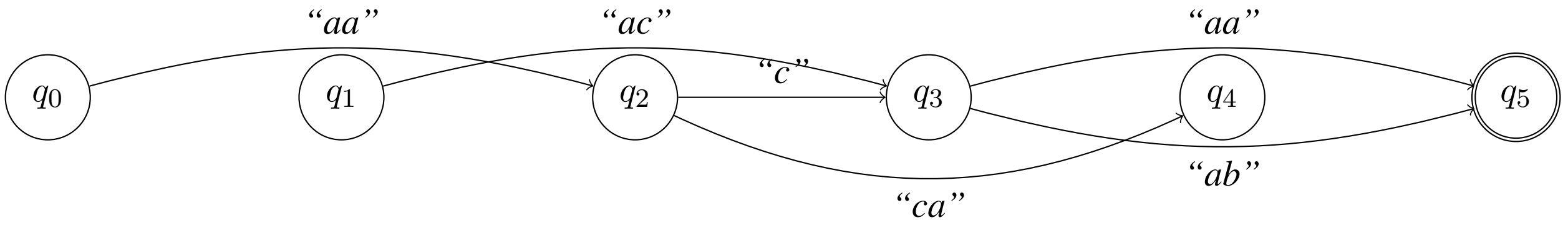

*For example, the word "aa" starting from $q_0$ yields the transitions $q_0 \xrightarrow{a} q_1 \xrightarrow{a} q_2$, creating edge $(q_0, \text{"aa"}, q_2)$. However, "aa" from $q_1$ fails because $q_2$ has no outgoing $a$-transition. The complete word-level automaton contains only those edges where the entire word can be successfully processed.*

*Finding grammatical paths. We seek paths from $q_0$ to $q_5$ using exactly three words (as required by grammar $G$). Examining the word-level automaton structure, we identify two valid paths:*
Path 1: $q_0 \xrightarrow{\text{"aa"}} q_2 \xrightarrow{\text{"c"}} q_3 \xrightarrow{\text{"aa"}} q_5$, *yielding sentence "aa c aa" with letter sequence "aacaa".*
Path 2: $q_0 \xrightarrow{\text{"aa"}} q_2 \xrightarrow{\text{"c"}} q_3 \xrightarrow{\text{"ab"}} q_5$, *yielding sentence "aa c ab" with letter sequence "aacab".*
*Both sentences are grammatically valid (each word belongs to $D$ and the sequence matches production $S \Rightarrow W\,W\,W$), and both letter sequences belong to $L_t$. The algorithm returns one such solution, for instance "aa c aa".*

*Therefore, this example admits the valid solutions "aa c aa" and "aa c ab". The algorithm returns one of them.*

## 6 Conclusion

We have established the decidability of the block-constrained sentence existence problem and provided polynomial-time algorithms for its solution. Our constructive approach, combining automata theory with context-free parsing, demonstrates that constrained text generation problems can be addressed through formal language theory. This work bridges historical combinatorial puzzles, formal language theory, and algorithmic design. By connecting circular partitions to grammatical constraints through generalized run-length encodings, we have shown how classical automata-theoretic techniques can be used to solve problems at the intersection of combinatorics and computational linguistics. These structured pattern constraints, as found in classical Arabic poetry and similar applications, are thus computationally tractable.

## Acknowledgments

The first two authors gratefully acknowledge support from the Global Scholar Award, University of Evansville Center for Innovation and Change. AI language models (ChatGPT-5.2 Instant and Claude Sonnet 4.5) were used occasionally in November and December 2025 to polish the manuscript, assist with figure generation via TikZ, and manage Arabic font handling. The authors also utilized these models to collaborate on developing and debugging Python programs for computational verification. All intellectual and scientific contributions are the authors' own work.